\documentclass[aps,twocolumn,groupedaddress,showpacs]{revtex4}
\usepackage[dvips]{graphicx}
\begin{document}
\title{Anomalies in the $ab$-plane resistivity of strongly underdoped
La$_{2-x}$Sr$_{x}$CuO$_{4}$ single crystals: possible charge
stripe ordering?}
\author{R.S.\@ Gonnelli}
\email[Corresponding author.E-mail:]{gonnelli@polito.it}%
\affiliation{INFM - Dipartimento di Fisica, Politecnico di
Torino, c.so Duca degli Abruzzi 24, 10129 Torino, Italy}
\author{V.A.\@ Stepanov}
%\email[]{stepanov@x4u.lpi.ruhep.ru}%
\affiliation{P.N. Lebedev Physical
Institute, Russian Academy of Sciences, SU-117924 Moscow, Russia}
\author{A.\@ Morello}
%\email[]{gonnelli@polito.it}%
\affiliation{INFM - Dipartimento di Fisica, Politecnico di
Torino, c.so Duca degli Abruzzi 24, 10129 Torino, Italy}
\author{G.A.\@ Ummarino}
%\email[]{ummarino@polito.it}%
\affiliation{INFM - Dipartimento di Fisica, Politecnico di
Torino, c.so Duca degli Abruzzi 24, 10129 Torino, Italy}
\author{D.\@ Daghero}
%\email[]{ddaghero@polito.it}%
\affiliation{INFM - Dipartimento di Fisica, Politecnico di
Torino, c.so Duca degli Abruzzi 24, 10129 Torino, Italy}
\author{L.\@ Natale}
%\email[]{luigi.natale@tin.it}%
\affiliation{INFM - Dipartimento di Fisica, Politecnico di
Torino, c.so Duca degli Abruzzi 24, 10129 Torino, Italy}
\author{Francesca Licci}
\affiliation{Istituto MASPEC del CNR, Parco Area delle Scienze,
37/A - 43010 Loc. Fontanini -  Parma, Italy}
\author{G.\@ Ubertalli}
\affiliation{Dipartimento di Scienza dei Materiali e Ing.\@
Chimica, Politecnico di Torino, 10129 Torino, Italy}
\date{\today}
%% ------------ Opzioni personali ------------
\newcommand{\ohm}{\ensuremath{{\rm\Omega}}}
\newcommand{\ped}[1]{\ensuremath{_{\rm #1}}}
\newcommand{\unit}[1]{\ensuremath{{\rm\,#1}}}
\newcommand{\gei}{\ensuremath{{\rm j}}}
\newcommand{\eu}{\ensuremath{{\rm e}}}
\newcommand{\micro}{\ensuremath{\mu}}
%% -------------------------------------------
\begin{abstract}
We present  and discuss the results of measurements of the
$ab$-plane resistivity of La$_{2-x}$Sr$_x$CuO$_4$ single crystals
with small Sr contents ($x=0.052\div 0.075$). The resistivity was
obtained in the temperature range between $4.2\unit{K}$ and
$300\unit{K}$ by using an AC Van der Pauw technique. When the
temperature is lowered, the $\rho\ped{ab}(T)$ curves show a
transition from a linear behaviour to a semiconducting-like one.
The deviation from the linearity occurs at a doping-dependent
temperature $T\ped{ch}$. This effect can be interpreted as due to
a progressive pinning of preexistent charge stripes. In the same
framework, the appearance of anomalous sharp peaks in the
$\rho\ped{ab}(T)$ curves of superconducting samples just above
$T\ped{c}$ may be explained by a transition from a \emph{nematic}
to a more ordered \emph{smectic} electronic stripe phase.
\end{abstract}
\maketitle
%%*************begin text*************************
One of the most debated features of doped perovskites is the
segregation of the charge carriers in one-dimensional domains,
called stripes, observed both in some superconducting samples
(i.e. Bi$_2$Sr$_2$CaCu$_2$O$_{8-\delta}$ \cite{Bianconi2}) and
more recently in non-superconducting samples \cite{Chen}. This
charge segregation is often invoked to explain the local survival
of the antiferromagnetic order typical of the undoped compounds,
but its correlation with the superconductivity is still a matter
of discussion. In the last few years many papers treating this
problem appeared in literature, stimulating the experimental
search for unquestionable evidences of stripe ordering in
high-$T\ped{c}$ cuprates.

Referring in particular to La$_{2-x}$Sr$_{x}$CuO$_{4}$ (LSCO) and
the related compound La$_{1.6-x}$Nd$_{0.4}$Sr$_{x}$CuO$_{4}$
obtained by Nd substitution, neutron diffraction and NQR
experiments \cite{Tranquada96,Imai99,Tranquada99b} allowed to
determine the temperatures of charge and spin ordering for
various Sr doping contents. In Ref.~\cite{Tranquada99b} a
signature of the charge ordering is identified in the deviation
of the $\rho_{\rm{ab}}(T)$ curve from the linearity which occurs
at a temperature $T\ped{ch}$. In principle, each phase transition
of the stripes involving a change in the transport properties
should leave a mark on the resistivity. Actually, if the metallic
character of the stripes \cite{Tranquada99b} is accepted, they
would be observable by means of resistivity measurements only
when the amplitude of their fluctuations is sufficiently small on
the measurement time scale, that is when they are at least
partially localized or ``pinned'' \cite{EmeryKivelsonTranquada}.
Therefore, $T\ped{ch}$ should be better referred to as a ``stripe
localization'' temperature.

In the case of La$_{2-x-y}$Nd$_y$Sr$_x$CuO$_4$, it can be argued
that the substitution of some La atoms with Nd stabilizes the
low-temperature tetragonal phase (LTT), which favours the stripe
pinning, instead of the low-temperature orthorhombic phase (LTO)
\cite{EmeryKivelsonTranquada,Tranquada99a}. Therefore the
occurrence of the LTO $\rightarrow$ LTT transition at about
$70\unit{K}$, besides leaving a clear kink on the resistivity
curves, gives rise to the pinning of preexistent stripes making
the $\rho_{\rm{ab}}(T)$ curve acquire a semiconducting-like
behaviour.

Actually, a similar transition from a metallic-like to a
semiconducting-like trend is also found in the $ab$-plane
resistivity of Nd-free samples with Sr contents $x=0.08$ and
$x=0.10$ \cite{Tranquada99b}. In these cases, however, a
different stripe pinning mechanism is expected, due to the
presence of impurities or to an intrinsic feature of
La$_{2-x}$Sr$_x$CuO$_4$.

In this paper, we report the results of measurements of the
$ab$-plane resistivity $\rho\ped{ab}(T)$ of strongly underdoped
LSCO crystals (\mbox{$0.052\leq x\leq 0.075$}) between 4.2 and
300~K. The behaviour of the measured $\rho\ped{ab}$ is then
discussed in terms of  three mechanisms following one another at
the lowering of the temperature: stripe localization, stripe
pinning and a possible stripe phase transition.

Platelet-like LSCO single crystals with mirror-like surfaces and
typical size $1\times 1\times 0.3\unit{mm}^{3}$ were grown by
slowly cooling down a non-stoichiometric melt. The high quality
of the crystals both from the superconductive and the
crystallographic point of view was shown by AC susceptibility
measurements and X-ray analysis. Energy Dispersion Spectroscopy
(EDS) microprobe analysis was used to determine the resulting
chemical composition of the samples, whose homogeneity was tested
by comparing the results of ``single-point'' analyses made in
different regions of each crystal.

The $ab$-plane resistivity was obtained by using an AC version of
the standard four-probe Van der Pauw technique. Four good
electrical contacts were obtained by evaporating Au on the
cleaved surface of each crystal, and then by submitting the
samples to an annealing process at about $900\unit{^{\circ}C}$ for
10 min in air. Thin gold wires were then soldered to the contacts
by using Ag conductive paint. Two of the leads were used to
inject into the samples alternating currents of $100\div
300\unit{\micro A}$ at $133\unit{Hz}$, while the remaining two
were connected to a lock-in amplifier. The $ab$-plane resistivity
was obtained by suitably averaging the results of \emph{four}
resistance measurements.

\begin{figure}[t]
\vspace{-3mm}
\includegraphics[keepaspectratio,width=8.6cm]{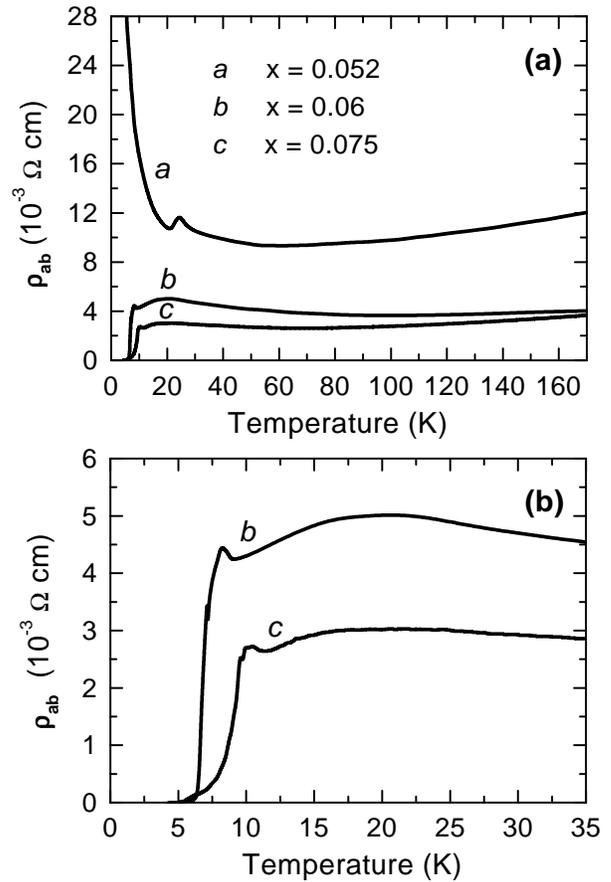}
\vspace{-10mm}\caption{\small{(a) The $ab$-plane resistivity of
our LSCO samples with $x = 0.052, 0.06$ and $0.075$ (curves
\emph{a}, \emph{b} and \emph{c}, respectively). (b) Enlargement of
the low-temperature part of curves \emph{b} and \emph{c} showing
the anomalous peaks above $T\ped{c}$.}} \vspace{-3mm}\label{f:1}
\end{figure}

Fig.~\ref{f:1}(a) shows the resulting $\rho\ped{ab}(T)$ curves for
samples with $x=0.052$ (curve~\emph{a}), 0.06 (curve~\emph{b})
and 0.075 (curve~\emph{c}). Crystals with $x=0.052$ do not undergo
a superconducting transition; instead, their resistivity shows a
steep growth as the temperature is lowered, in a typical
semiconducting-like fashion. However, an evident well-reproducible
anomalous maximum followed by a dip is observed at a temperature
of about $25\unit{K}$. We reserve to give a detailed discussion
of these features in a future paper.

Samples with $x=0.06$ and $x=0.075$ have instead a
superconducting behaviour. The transition of the crystals with
$x=0.06$ occurs at a critical temperature (evaluated at the
midpoint of the transition) $T\ped{c} = 6.7\unit{K}$ and has a
broadening $\Delta T\ped{c} \rm{(10\div90\%)} = 0.6\unit{K}$. For
the crystals having $x=0.075$ the analogous quantities are
$T\ped{c} = 8.9\unit{K}$ and $\Delta T\ped{c} = 2.2\unit{K}$. As
shown more clearly in Fig.~\ref{f:1}(b), in both these cases the
resistivity shows a sharp anomalous peak of width $\Delta
T\approx 1 \unit{K}$ at a temperature $T\ped{so}$ about
$1.5\unit{K}$ greater than the respective $T\ped{c}$.
Incidentally, no anomalies are present in the AC susceptibility
curves of the same samples.

Before giving an interpretation to the observed behaviours of the
$ab$-plane resistivity curves, it is worthwhile to mention and
confute some possible objections to the results obtained.

First, one could wonder if the peaks in curves \emph{b} and
\emph{c} of Fig.~\ref{f:1}(b) could be produced by some inaccuracy
in the measurement or by an intrinsic systematic error due to the
application of the Van der Pauw method. However, these curves are
very reproducible and the experimental errors are not greater
than the thickness of the lines. Moreover, the anomalous peak is
present and well recognizable in \emph{all} the resistance vs.\@
temperature curves (with more than 2000 points each) which have
been averaged to obtain the $ab$-plane resistivity as required by
the Van der Pauw measurement technique, and therefore it is not
likely to be an artifact introduced by inhomogeneities in the
samples.

\begin{figure}[t]
\vspace{-3mm}
\includegraphics[keepaspectratio,width=8.6cm]{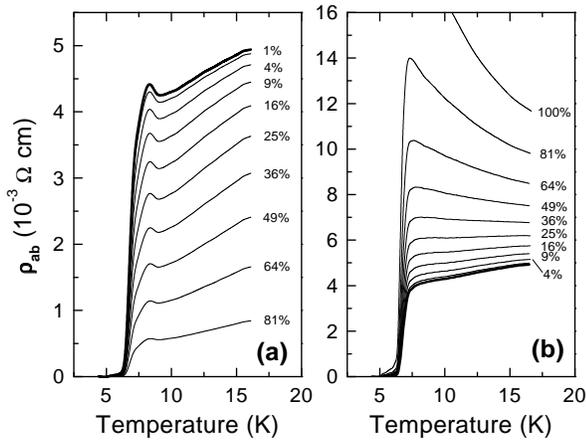}
\vspace{-3mm}\caption{\small{(a) Resistivity curves obtained by
removing the contribution of increasing volume percentage of
non-superconducting inhomogeneity from the measured
$\rho\ped{ab}(T)$ curve (bold line) of the LSCO samples with $x =%
0.06$.  (b) $\rho\ped{ab}(T)$ curves for $x =%
0.06$ allowing for $c$-axis contribution, calculated for the
resistor network described in the text. The bold line is the
measured $\rho\ped{ab}(T)$ without the peak at $T\ped{so}$. }}
\vspace{-3mm} \label{f:2}
\end{figure}

Another possible objection is that the peaks in the curves
\emph{b} and \emph{c} could be due to the contribution of
semiconducting-like,  non-superconducting regions of the sample.
To check whether this is the case, we removed from the measured
in-plane resistivity given by curves \emph{b} and \emph{c} of
Fig.~\ref{f:1}(b) the contribution of non-superconducting regions
of different size, representing low-doping islands whose $\rho
(T)$ is given by curve~\emph{a} of the same figure. The results
reported in Fig.~\ref{f:2}(a) for various non-superconducting
volume percentage show that there's no way to cancel the
anomalous peak above $T\ped{c}$.

Finally, one could argue that a contribution of the $c$-axis
resistivity $\rho\ped{c}$ (which increases rapidly as
$T\rightarrow 0$) could affect the measured $\rho\ped{ab}$ giving
rise to a peak such as that observed. Actually, it is important to
note that: i) $\rho\ped{c}$ is, at low temperatures, about 8000
times $\rho\ped{ab}$ \cite{Boebinger}; ii) because of the
annealing process, the gold electrodes deeply penetrate into our
samples, thus connecting several consecutive \emph{ab} planes. If
there were no inhomogeneities in the crystal, the current would
only flow along these $ab$ planes disregarding the \emph{c}-axis
paths. Consequently, a $c$-axis contribution is expected only in
the presence of inhomogeneities in the $ab$ planes. In order to
analyze what happens in this case, we numerically calculated the
equivalent resistance of a resistor network, in which each
elementary cell represents the unit volume of the crystal (with
vertical and horizontal resistances numerically equal
respectively to the measured $\rho\ped{c}$ and $\rho\ped{ab}$ of
our samples after the removal of the anomalous peak from
$\rho\ped{ab}$), by assigning to a certain number of horizontal
resistor the non-superconducting $\rho (T)$ given by
curve~\emph{a} in Fig~\ref{f:1}(a). The results, shown in
Fig.~\ref{f:2}(b) for increasing non-superconducting volume
percentage, indicate that the $c$-axis contribution can strongly
modify the shape of the $\rho(T)$ curve, but is unable to create
very sharp peaks as those observed in our measurements.

\begin{figure}[t]
\vspace{-3mm}
\includegraphics[keepaspectratio,width=8.6cm]{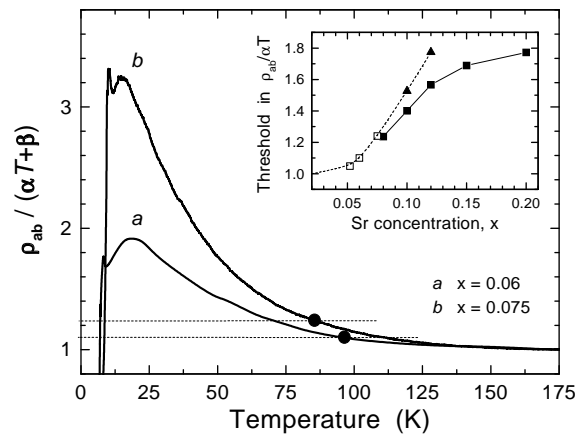}
\vspace{-3mm}\caption{\small{Normalized $ab$-plane resistivity
curves for the samples with $x=0.06$ (curve a) and $x=0.075$
(curve b). The curve for $x=0.052$ is not reported for clarity.
The inset shows
the thresholds in $\rho\ped{ab} (T)/(\alpha\ped{0} T +%
\beta\ped{0})$ as functions of the Sr doping in Nd-doped
\cite{Tranquada99b} and Nd-free samples (Ref.~\cite{Tranquada99b}
and our extrapolation). For details see the text.}} \vspace{-3mm}
\label{f:3}
\end{figure}

We can now discuss in detail the results shown in Fig~\ref{f:1}(a)
and (b). In the range \mbox{$150\leq T\leq 300\unit{K}$} the
$ab$-plane resistivity of all the crystals considered has an
almost perfectly linear temperature dependence. We then
perform a linear fit with a curve of the form \mbox{$\alpha%
T + \beta$} which yields the best-fit parameters $\alpha\ped{0}$
and $\beta\ped{0}$. With these values we calculate the normalized
$ab$-plane resistivity $\rho\ped{ab} (T)/(\alpha\ped{0} T +%
\beta\ped{0})$ in the \emph{whole} temperature range, from 4.2 up
to 300$\unit{K}$. The resulting curves for the superconducting
samples are plotted in Fig.~\ref{f:3}.

\begin{figure}[t]
\vspace{-3mm}
\includegraphics[keepaspectratio,width=8.6cm]{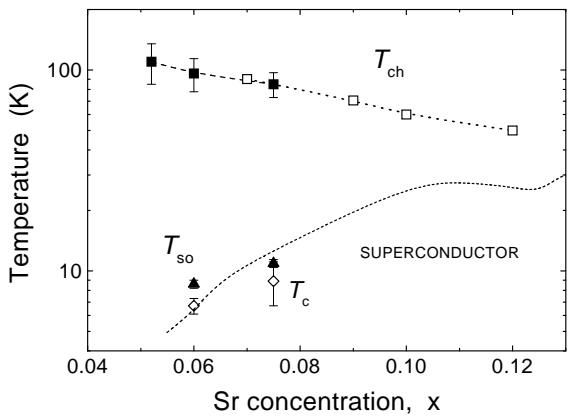}
\vspace{-6mm}\caption{\small{Stripe localization temperatures
$T\ped{ch}$ (solid squares), temperatures of the observed
anomalous peaks in the $ab$-plane resistivity $T\ped{so}$ (solid
triangles) and critical temperatures $T\ped{c}$ (open diamonds)
determined from our resistivity data together with $T\ped{ch}$
from NQR data \cite{Imai99,Singer} (open squares) as functions of
the doping (for details see the text).}} \vspace{-3mm} \label{f:4}
\end{figure}

As previously anticipated, in a recent paper Ichikawa \emph{et
al.}\@ \cite{Tranquada99b} noted that the stripe localization
temperature $T\ped{ch}$ (previously determined by NQR
\cite{Imai99,Singer}) of La$_{2-x}$Sr$_x$CuO$_4$ single crystals
with $x=0.10$ and $x=0.12$ marks the point at which the
corresponding $ab$-plane resistivity deviates from the
high-temperature linear behaviour of an amount that depends on
the Sr concentration. In such a way, they defined threshold
values for the deviation from linearity, which allow in principle
to determine $T\ped{ch}$ once known $\rho\ped{ab}(T)$. Supported
by their work, we extrapolated the threshold vs.\@ $x$ curve
given in Ref. \cite{Tranquada99b} to the Sr contents of our
interest: $x=0.052$, 0.06 and 0.075 (see the inset of
Fig.~\ref{f:3}). The extrapolation procedure is based on the fact
that, for small $x$, the threshold must tend to 1 and on a
somehow arbitrary choice for the extrapolating function. By using
B-spline functions we determined the smoothest curve that
connects the straight line passing for the points deduced from
Ref. \cite{Tranquada99b} (solid triangles in the inset of
Fig.~\ref{f:3}) to the horizontal line of equation $y=1$, where
$y$ is the threshold value. To do so, we used as an additional
point the intersection between these two straight lines, and we
put to 1 the threshold at $x\leq0.02$ by using the fact that, for
these doping contents, the material is an antiferromagnetic
insulator. The thresholds determined in such a way for our Sr
concentrations are reported as open squares in the inset of
Fig.~\ref{f:3} together with the thresholds for Nd-doped crystals
reported in Ref.\@ \cite{Tranquada99b} (solid squares). The point
at which the normalized resistivity crosses the threshold
determines the ``stripe localization'' temperature $T\ped{ch}$
for each doping concentration. The crossing points for $x=0.06$
and $x=0.075$ are reported in the main graph of Fig.~\ref{f:3}
(solid circles). The charge stripe localization temperatures
$T\ped{ch}$ are then plotted as a function of the Sr content in
Fig.~\ref{f:4} (solid squares) together with the data (open
squares) obtained by NQR measurements on La$_{2-x}$Sr$_x$CuO$_4$
crystals with various doping contents \cite{Imai99,Singer}.
There's a \emph{very good agreement} between the two data sets,
which seems to indicate that the procedure described so far is
essentially correct.

As shown in Fig.~\ref{f:3}, when the temperature is lowered below
$T\ped{ch}$ the normalized $\rho\ped{ab}$ increases, until the
superconducting transition sets on. The upturn of the curve can
be related to a progressive slowing down of the stripe
fluctuations, i.e.\@ to an increase of the pinning efficiency.
Actually, the problem arises of determining what is the mechanism
responsible for the stripe pinning, since there is no Nd
substitution nor evidences of a structural lattice transition
which would perhaps result in a kink in the $\rho\ped{ab}$. We
propose that, in the strongly underdoped crystals we analyzed,
the stripe pinning is ``intrinsic'', in the sense that it is
related to the very low density of doped holes ($1.052\div 1.075$
holes per Cu site).

Let us now focus our attention on the narrow peaks observed in the
superconducting samples ($x=0.06$ and $x=0.075$) at $T\ped{so}$
(solid triangles in Fig.~\ref{f:4}) just above $T\ped{c}$ (open
diamonds). A possible explanation for these anomalies could be
the occurrence of a \emph{structural} lattice transition, of the
same kind of that observed in Nd-substituted samples at higher
temperature \cite{Tranquada99b}. Actually, to our knowledge there
are no other kinds of experiments supporting the occurrence of
such a transition near $T\ped{c}$ in La$_{2-x}$Sr$_x$CuO$_4$.
Thus, we think these peaks could be due to an \emph{electronic}
transition to a more ordered stripe phase. This argument is based
on Ref.~\cite{Emery}, in which a detailed scenario of successive
electronic phase transitions is depicted. As suggested by the
schematic phase diagram reported there, a \emph{nematic} to
\emph{smectic} phase transition just above the critical
temperature is allowed in the case of underdoped (but obviously
superconducting) copper-oxides. Such a transition is not in
conflict with the stripe localization argument exposed above.
When the \emph{progressive} intrinsic pinning of the stripes
begins, it is possible that transverse stripe fluctuations are
sufficiently violent to preserve the full translational
invariance, destroying the rotational symmetry, as in the
\emph{nematic} stripe phase \cite{Emery}. At the lowering of the
temperature, transverse fluctuations decrease and the stripes can
undergo the transition to a \emph{smectic} phase, or rather to a
``stripe glass'', which is in fact a slightly disordered version
of the smectic phase~\cite{Emery}.

The possibility of observing the mark of an electronic phase
transition on the resistivity is also supported by experimental
analogies with the behaviour of some metal dichalcogenides at the
charge density wave transition \cite{dichalcogenides}. There,
clear kinks very similar to the features present in our data were
observed in the $\rho\ped{ab}(T)$ curves.

The interpretation of the anomalous peaks at $T\ped{so}$ exposed
above is also consistent with the coexistence of cluster
spin-glass (CSG) and superconductivity recently observed
below~$\sim 5\unit{K}$ in LSCO with $x=0.06$ \cite{Julien}. In
fact, as predicted by the general theory of coupled order
parameters, a spin order can appear only when the charge order is
already well developed, i.e.\@ at a lower temperature
\cite{Emery}. The temperature of stripe ordering we determined
here ($T\ped{so} = 8.6$ K) is, as expected, greater than the
temperature at which the CSG phase appears ($T\ped{csg} \sim
5\unit{K}$).

In conclusion, we interpret the main features of the $ab$-plane
resistivity of strongly underdoped La$_{2-x}$Sr$_x$CuO$_4$
crystals as a function of the temperature within the framework of
the stripe phase transitions. We argue that the deviation from the
linearity of the normalized $\rho\ped{ab}(T)$ curves allow
obtaining the temperature of stripe localization $T\ped{ch}$. At
the lowering of the temperature below $T\ped{ch}$, we hypothesize
that an intrinsic stripe pinning mechanism due to the very low Sr
content of our samples becomes dominant giving rise to the upturn
of the $\rho\ped{ab}(T)$. Finally, we propose to explain the
anomalous narrow peaks above $T\ped{c}$ as due to a nematic to
smectic electronic phase transition of the stripes. \vspace{-8mm}
\\

\section*{ACKNOWLEDGEMENTS} \vspace{-3mm}
Many thanks are due to A.~Rigamonti for the useful discussions.
This work has been done under the Advanced Research Project SPIS
of the Istituto Nazionale per la Fisica della Materia (INFM).

% ------------------------------------------------------------------
%\bibliography{stripes}
%
\end{document}